\documentclass[preprint,superscriptaddress, amsmath,amssymb, aps, floatfix]{revtex4-2}
\usepackage{geometry}
\usepackage{textcomp, gensymb}
\usepackage{amsmath}
\usepackage{xcolor}
\usepackage{graphicx}
\usepackage{dcolumn}
\usepackage{bm}
\usepackage{hyperref}
\usepackage{siunitx}
\hypersetup{colorlinks = true,linkcolor=blue,
     filecolor=blue,
     urlcolor=blue,
     citecolor = blue, pdfauthor=author}
\begin{document}
\renewcommand{\figureautorefname}{Fig.}

\title{W-shaped Broadband Attenuation of Longitudinal Waves through Composite Elastic Metamaterial}

\author{Brahim Lemkalli}
\email{\textcolor{black}{brahim.lemkalli@femto-st.fr}}
\affiliation{Universit\'{e} Marie et Louis Pasteur, Institut FEMTO-ST, CNRS, 25000 Besan\c{c}on, France}
\author{Krzysztof K. Dudek}
\affiliation{Universit\'{e} Marie et Louis Pasteur, Institut FEMTO-ST, CNRS, 25000 Besan\c{c}on, France}
\affiliation{Institute of Physics, University of Zielona Gora, ul. Szafrana 4a, Zielona Gora 65-069, Poland}
\author{Muamer Kadic}
\affiliation{Universit\'{e} Marie et Louis Pasteur, Institut FEMTO-ST, CNRS, 25000 Besan\c{c}on, France}
\author{Qingxiang Ji}
\affiliation{Universit\'{e} Marie et Louis Pasteur, Institut FEMTO-ST, CNRS, 25000 Besan\c{c}on, France}
\author{S\'{e}bastien Guenneau}
\affiliation{The Blackett Laboratory, Physics Department, Imperial College London, SW7~2AZ, London, UK}
\affiliation{UMI 2004 Abraham de Moivre-CNRS, Imperial College London, SW7~2AZ, London, UK}

\author{Abdellah Mir}
\affiliation{Department of Physics, Moulay Ismail University, B.P. 11201, Zitoune, Meknes, Morocco}
\author{Younes Achaoui}
\affiliation{Laboratory of Optics, Information Processing, Mechanics, Energetics and Electronics, Department of Physics, Moulay Ismail University, B.P. 11201, Zitoune, Meknes, Morocco}
\date{\today}
\begin{abstract}
We investigate a composite elastic meta-slab with exceptional transmission properties, particularly the presence of a W-shaped bandgap. A comprehensive study, utilizing experimental measurements, the finite element method, and an analytical approach, identifies this specific bandgap. The meta-slab design involves cutting an array of composite materials arranged in parallel with strategically placed incisions. This configuration ensures that the materials between the slits act as plate-like waveguides within the surrounding medium. The incorporation of steel into ABS-based Fabry-Perot cavities induces a notable coupling effect between longitudinal waves and localized modes traversing the structure, leading to the formation of two distinct Fabry-Perot resonators. These coupling effects generate a series of resonances and antiresonances, ultimately producing the W-band gap through the interaction of two symmetric Fano resonances.
\end{abstract}

\maketitle

\section{Introduction}

Phononic crystals (PnCs) and elastic metamaterials are two areas of emerging research that have sparked interest in the context of broadband wave propagation issues in recent years \cite{krushynska2023emerging, lu2009phononic,vasileiadis2021progress, hu2021acoustic}. On one hand, PnCs are scatterer arrangements that have been specifically structured within a particular pattern \cite{khelif2015phononic, pennec2010two}. These complex arrangements produce band gaps due to a Bragg scattering mechanism in which the wavelength of the incident waves matches the periodicity parameters of the crystal's lattice \cite{pennec2016fundamental, achaoui2011experimental}. Besides, obtaining low-frequency band gaps in PnCs often requires a wide array of scatterers, demonstrating the significant challenges involved in controlling waves at these scales. On the other hand, the year 2000 was a turning point with the introduction of the local resonance \cite{liu2000locally}. This occurrence paved the way for lower-frequency bandgap engineering without the restrictions imposed by periodic resonator arrangements \cite{ma2016acoustic, krushynska2017coupling}. The key point of this realization is that even sizes smaller than the wavelength of propagating waves can now control their behavior, opening the way to an approach to wave management that goes beyond PnCs limits. The local resonance paradigm is based on the role of resonators and their extraordinary ability to control wave dynamics. These resonators, when properly designed, possess the unique ability to induce band gaps even in the absence of complete periodicity \cite{zaky2023design}. Based on the fundamentals of local resonance, the concept of metamaterials has become a major focus of attention, providing a perfect illustration of the ability of these artificial structures to take on their own unique characteristics \cite{kadic20193d}. Metamaterials, whether they conform to periodic or aperiodic structures, depart from the limits of conventional materials \cite{lemkalli2023mapping}. One of the defining features of metamaterials is their ability to manipulate waves - whether electromagnetic \cite{zheludev2012metamaterials, zhou2010level, sihvola2007metamaterials}, acoustic \cite{cummer2016controlling, guenneau2007acoustic, liao2021acoustic} or elastodynamic \cite{huang2021recent, oh2016elastic, bigoni2013elastic}- in ways that break with traditional norms. Through careful manipulation of the geometry, size and arrangement of unit cells at sub-wavelength scales, metamaterials open the way to a realm of extraordinary functionalities. 

In general, band gaps are required for numerous applications, including noise reduction \cite{gao2022acoustic, ma2021structural}, vibration shielding \cite{wu2020mechanical, miniaci2016large, luo2023surface}, sensing \cite{zaremanesh2021temperature, alrowaili2023heavy, almawgani2023periodic, lucklum2012two}, and filtering \cite{pennec2004tunable, wu2008narrow,lemkalli2024longitudinal, ji2016development, qiu2005mode, chen2017acoustic}, all of which are very promising areas of research. Filtering and sensing are encompassed in the fundamental process, which is the emergence of defects in the main arrangements in order to generate a certain localized resonance mode within the band gap \cite{wang2022brief, valipour2022metamaterials, yan2022band}. These are most commonly studied in phononic crystal, which in today's technology-driven world employ a whole range of technologies and designs \cite{lucklum2009phononic, gharibi2019very, lucklum2021phononic}. These applications have introduced distinct and versatile types of control devices capable of operating in a wide range of frequencies by leveraging the phenomenon of local resonance, also known as Fano resonances \cite{zaki2020fano, oudich2018rayleigh, wang2020robust}. Also, it is often difficult to achieve a broadening of the band gap in the low frequency range as well as the presence of a band gap at these frequencies. In order to overcome these challenges, a variety of metamaterials has been developed, including elastic metamaterials with nonlinear damping \cite{zhang2025borrow}, mechanical metamaterials with negative stiffness \cite{chen2023elastic}, three-dimensional nested hybrid lattice structures \cite{li2024mechanisms}, and bioinspired architected metamaterials \cite{li2024unprecedented}. In addition to this, several researches have emphasized different characteristics resulting from the utilization of Fabry-Perot cavities to generate band gaps. Elayouch et al. studied the transmission properties of an acoustic metamaterial composed of two Fabry-Perot cavities of the type space-coiled cavities. These metamaterials operate at low frequency through the generation of a series of resonances and antiresonances, leading to an attenuation band gap for sound waves \cite{elayouch2017subwavelength}. More recently, Sellami et al. demonstrated an adjustable phononic membrane with Fabry-Perot microcavities that operates as a temperature sensor. The phononic membrane is perforated with subwavelength slits and emerges in water, showing a band gap for ultrasonic waves with a relative bandwidth of 31$\%$ \cite{sellami2023experimental}. This leads to a quasi-bound state in the continuum for ultrasonics waves, offering a framework for the design of efficient, ultra-high Q-factor ultrasound devices \cite{farhat2024observation}.

\begin{figure}[!h]
    \centering
    \includegraphics[width=1\linewidth]{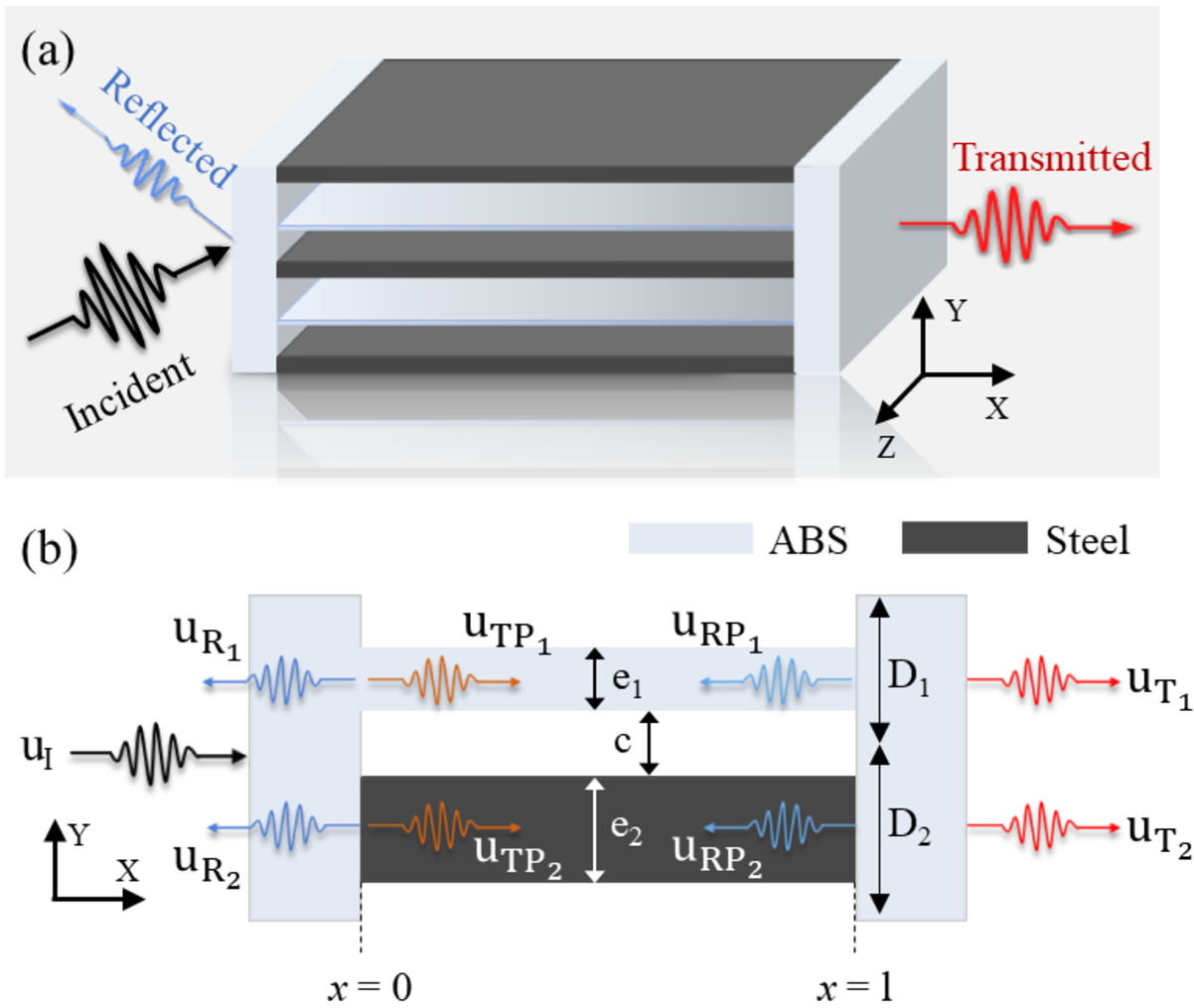}
    \caption{The composite elastic metamaterial. (a) Schematic of the metamaterials which consists of rectangular slots of length $l=6a$ and width $c=0.225a$, separated by ABS of width $e_1=0.15a$ and steel of width $e_2=0.4a$. Here $a$ represents the lateral periodicity. The incident longitudinal wave (black), total reflected wave (blue), and total transmitted wave (red) are also indicated. (b) A zoomed-in view of the unit cell illustrates the incident wave (with displacement $\rm u_I$) and the two transmitted waves (with displacement $\rm u_{{\rm T}_1}$ and $\rm u_{{\rm T}_2}$).
    We also mark the local reflected waves at the interface ABS-ABS ($\rm u_{{\rm R}_1}$ and $\rm u_{RP_1}$) and interface ABS-Steel ($\rm u_{{\rm R}_2}$ and $\rm u_{{\rm RP}_2}$), respectively.}
    \label{figure 1}
\end{figure}
In this work, we examine in depth the W-shaped band gap created by a composite elastic metamaterial. The objective of our study is to figure out how this band gap in the structure operates. The design of the metamaterial is made up of a single unit cell that is aligned with the opposite direction of wave propagation, as depicted in \autoref{figure 1}. The transmission curve is used to validate the accuracy of our band gap calculations and experimental measurement. Also, the causes of the band gap are explained via dispersion analysis, which relates to an infinite structure scenario, a context separate from the one we are working with. On top of that, we investigate the coupling phenomenon between two resonators, which generates the W-shaped band gap at lower frequencies. Our research also looks into the effect of size changes on the band gap and its relationship to the diffraction limit.

\section{Materials and methods}
\subsection{Numerical models}

The elastic metamaterial that we propose consists of two rectangular slits, each with a length $l$ and a width $c$, in a homogeneous material known as Acrylonitrile Butadiene Styrene (ABS). These slits are separated by a steel bar with a width $e_2$. Such arrangement gives rise to an array of decoupling resonators situated on the side perpendicular to the propagation direction of longitudinal waves. Essentially, this configuration forms local mechanical resonators in the $xy$-plane, achieved by embedding steel elements within an ABS panel with rectangular holes in between. A partial depiction of this proposed elastic metamaterial, along with its unit cell, is illustrated in \autoref{figure 1}. We should note that the use of slits in a homogeneous material has precedents proposed in several elastic metamaterials for focusing \cite{su2016focusing}, splitting shear vertical and pressure waves \cite{su2018elastic} and reflecting \cite{ruan2021reflective} many elastic waves. The distinction between these efforts is that we use a composite elastic metamaterial to create a longitudinal wave band gap in the W-shape, which is an unusual concept in elastic metamaterials.

We study the elastodynamic behaviors numerically by the commercial software COMSOL Multiphysics, which solves the Navier equations in their weak forms, as represented by:
\begin{equation}\label{Eq01}
    \frac{E}{2(1+\nu)}\biggl(\frac{1}{(1-2\nu)}\nabla (\nabla.\textbf{u})+\nabla^2 \textbf{u}\biggl)=-\omega^2\rho\textbf{u},
\end{equation}
where $\textbf{u}(x, y, z)$ is the displacement vector, $E$, $\nu$ are the Young's modulus and Poisson's ratio, respectively, $\omega$ is the angular wave frequency and $\nabla$ is the gradient. In the computational model, we define all materials as isotropic linear elastic materials, with their mechanical properties shown in \autoref{Table1}. For the meshing, we took into account a maximum element size of $\lambda_{\text{min}}/10$ and tetrahedral elements. We then integrated the displacement for both the input and the output media in order to calculate the transmission.

The transmission of the longitudinal wave is calculated by:
\begin{equation}\label{Eq02}
{\rm Transmision} \;{(\rm dB)}=20\log_{10}\left(\frac{\textbf{u}_{\rm T}(x)}{\textbf{u}_{\rm I}(x)}\right),
\end{equation}
where $u_{T}(x)$ and $u_{\rm I}(x)$ are the displacement vectors of the incident and transmitted longitudinal waves in the $x$-direction, respectively.

\begin{table}[!h]
    \centering
    \setlength{\tabcolsep}{0.2cm}
    \caption{ \label{Table1} The material properties.}
    \begin{tabular}{c c c c}\hline
     Materials & Young's modulus & Poisson's ratio & Density   \\
      & (\si{GPa}) & & (\si{kg/m^3})\\\hline\\
     Steel & 201 & 0.33 & 7843\\ 
     ABS & 2.6 & 0.4 & 1020\\\hline
    \end{tabular}
\end{table} 

Following our examination on the W-shaped band gap transmission in the elastic metamaterial, we proceed to explore dispersion characteristics by conducting simulations involving an eigenvalue problem. In this analysis, we compute the phononic dispersion relations, taking into account the periodic repetition of cells along the $x$-direction and the infinite extension in the $y$-direction. To ensure accuracy in our simulations, we implement Floquet-Bloch boundary conditions along the $x$-direction , which are applied as follows:

\begin{equation}\label{eq03}
    \textbf{u}(x+d)=\textbf{u}(x)e^{ik_xd},
\end{equation}
where $d=l+0.6a$ is the array pitch, and ${\bf k}=(k_x, 0, 0)$ is the reduced Bloch wavevector and $\textbf{u}(x)$ is the displacement vector in $x$-direction.

We illustrate the dispersion relationship within the initial irreducible Brillouin zone, denoted as $\Gamma \rm X$. Here $\Gamma$ corresponds to the point ($0$, $0$, $0$) and $\rm X$ represents ($\pi/d$, $0$, $0$). In order to characterize the polarization of longitudinal waves, we compute the dispersion curves while considering polarization weighting, employing the following expression:

\begin{equation}\label{eq04}
    p_{x}=\frac{\iint_S |\textbf{u}(x)|\; dxdy}{\iint_S \sqrt{|\textbf{u}(x)|^2+ |\textbf{u}(y)|^2}\; dxdy},
\end{equation}
where $S$ is the total surface of the unit cell, $|\textbf{u}(x)|=\sqrt{\textbf{u}(x)~ \textbf{u}^*(x)}$ is the modulus of displacement vector, with $\textbf{u}^*(x)$ is the complex conjugate of $\textbf{u}(x)$. 

\subsection{Analytical model}
Our approach to developing an analytical model for describing the transmission of longitudinal waves through elastic metasurface draws from well-established theoretical approaches outlined in references \cite{su2018elastic, su2016focusing}. This model's construction is relatively straightforward, as illustrated in \autoref{figure 1}(b), where the diagram depicts the incident longitudinal wave alongside its reflected and transmitted counterparts. To formulate our analytical model, we integrate the definitions and continuity conditions mentioned above for the presented design. In order to construct an analytical framework capable of elucidating the existence and characteristics of the W-shaped band gap within our system.

We consider the configuration depicted in \autoref{figure 1}(b), which shows another perspective of the unit cell. This unit cell consists of an array of plates made from ABS and Steel materials, and it is situated between two free media also composed of ABS. This configuration effectively creates two homogeneous solids with aligned rectangular slits. At each of these junctions between the plates and the free media, there is an interaction between the incident longitudinal wave in the left-side free medium and the longitudinal waves within the plates.

In essence, when an incident longitudinal wave enters the first free medium from the left, it propagates through the plates as a longitudinal wave with speed: $C_{L-Pi}=\sqrt{\frac{E_i}{\rho_i(1-\nu_i)}}$, where $i$ symbolizes ABS or steel material. Subsequently, it is transmitted to the right side and reverts back into a longitudinal wave upon entering the other free medium, which is propagated in both free ABS media with the same speed: $C_{L}=\sqrt{\frac{E(1-\nu)}{\rho (1+\nu)(1-2\nu)}}$. We should note that the displacements in each direction are denoted as $u(x)$, $u(y)$, and $u(z)$, respectively. We make the assumption that the incident plane wave is independent of the $y$ and $z$-directions, implying there are no $y$ and $z$-dependent terms.

To calculate the transmission of longitudinal waves, we can represent the amplitudes of the displacement as follows: $I$ for the incident wave, $R$ for the reflected wave, and $T$ for the transmitted wave. Additionally, we denote the amplitudes of displacement on the ABS plate as $A_1$ and $B_1$, and for the Steel plate, we use $A_2$ and $B_2$. The displacements can then be expressed as:

\begin{equation}\label{Eq05}
    u(x<0)=I e^{i(k_{L}x-\omega t)}+R_1 e^{-i(k_{L}x+\omega t)}+R_2 e^{-i(k_{L}x+\omega t)},
\end{equation}
\begin{multline}\label{Eq06}
    u(0<x<l)=A_1 e^{i(k_{L-P1}x-\omega t)}+B_1 e^{-i(k_{L-P1}x+\omega t)}+A_2 e^{i(k_{L-P2}x-\omega t)}+B_2 e^{-i(k_{L-P2}x+\omega t)},
\end{multline}
\begin{equation}\label{Eq07}
    u(l<x)=T_1 e^{i(k_{L}x-\omega t)}+T_1 e^{i(k_{L}x-\omega t)},
\end{equation}
where $k_{L}$ is the wavenumber of the longitudinal wave in the free ABS media, $k_{L-P1}$ is the wavenumber of the longitudinal wave in ABS plate, and $k_{L-P2}$ is the wavenumber of the longitudinal wave in Steel plate.

By applying the continuity conditions of the displacements in the $x$-direction and the compressional forces $F_x=E_{ABS}D_i\frac{\partial u(x)}{\partial x}$ in the free medium and $F_x=E_ie_i\frac{\partial u(x)}{\partial x}$ in the plates at both $x=0$ and $x=l$, we found the following equations:
\begin{equation}\label{Eq08}
   I+R_1+R_2=A_1+B_1+A_2+B_2,
\end{equation}
\begin{equation}\label{Eq09}
   I-R_1-R_2=\alpha(A_1-B_1)+\beta(A_2-B_2),
\end{equation}
\begin{equation}\label{Eq10}
   (T_1+T_2)e^{ik_{L}l}=A_1e^{ik_{L-P1}l}+B_1e^{-ik_{L-P1}l}+A_2e^{ik_{L-P2}l}+B_2e^{-ik_{L-P2}l},
\end{equation}
\begin{equation}\label{Eq11}
   (T_1+T_2)e^{ik_{L}l}=\alpha(A_1e^{ik_{L-P1}l}-B_1e^{-ik_{L-P1}l})+\beta(A_2e^{ik_{L-P2}l}-B_2e^{-ik_{L-P2}l}).
\end{equation}
where $\alpha =\frac{e_1}{D_1+D_2}\frac{k_{L-P1}}{k_{L}}$ and $\beta =\frac{E_{ABS}}{E_{Steel}}\frac{e_2}{D_1+D_2}\frac{k_{L-P2}}{k_{L}}$.

By solving the above equations, we find that the expression of the transmission coefficient as a function of the characteristics of both plates for the case of an incident longitudinal wave with an amplitude of $I = 1$, a reflected wave with and amplitude $R=R_1+R_2$, and a transmitted wave with and amplitude $T=T_1+T_2$ is then given by:
 
\begin{equation}\label{Eq12}
   T=\frac{(\alpha k_{L-P1}+\beta k_{L-P2})4k_{L}e^{i\frac{k_{L-P1}+k_{L-P2}}{2}l}}{f_1^2-f_2^2 e^{i2k_{L-P2}l}-f_3^2 e^{i2k_{L-P1}l}+f_4^2 e^{i2l(k_{L-P1}+k_{L-P2})}},
\end{equation}
with
\begin{equation}\label{Eq13}
   f_1=k_{L}^2+\beta k_{L}k_{L-P2}+ \alpha k_{L}k_{L-P1}+\alpha \beta k_{L-P1}k_{L-P2},
\end{equation}
\begin{equation}\label{Eq14}
   f_2=k_{L}^2-\beta k_{L}k_{L-P2}+ \alpha k_{L}k_{L-P1}-\alpha \beta k_{L-P1}k_{L-P2},
\end{equation}
\begin{equation}\label{Eq15}
   f_3=k_{L}^2+\beta k_{L}k_{L-P2}- \alpha k_{L}k_{L-P1}-\alpha \beta k_{L-P1}k_{L-P2},
\end{equation}
\begin{equation}\label{Eq16}
   f_4=k_{L}^2-\beta k_{L}k_{L-P2}- \alpha k_{L}k_{L-P1}+\alpha \beta k_{L-P1}k_{L-P2}.
\end{equation}

To compare the analytical results with those obtained using the finite element approach, we utilize the transmission in decibels (dB), which is written as follows:
\begin{equation}\label{Eq17}
   \rm Transmission( \rm dB)=20log_{10}(T).
\end{equation}

\subsection{Experimental validation}

\begin{figure}[!h]
    \centering
    \includegraphics[width=9cm]{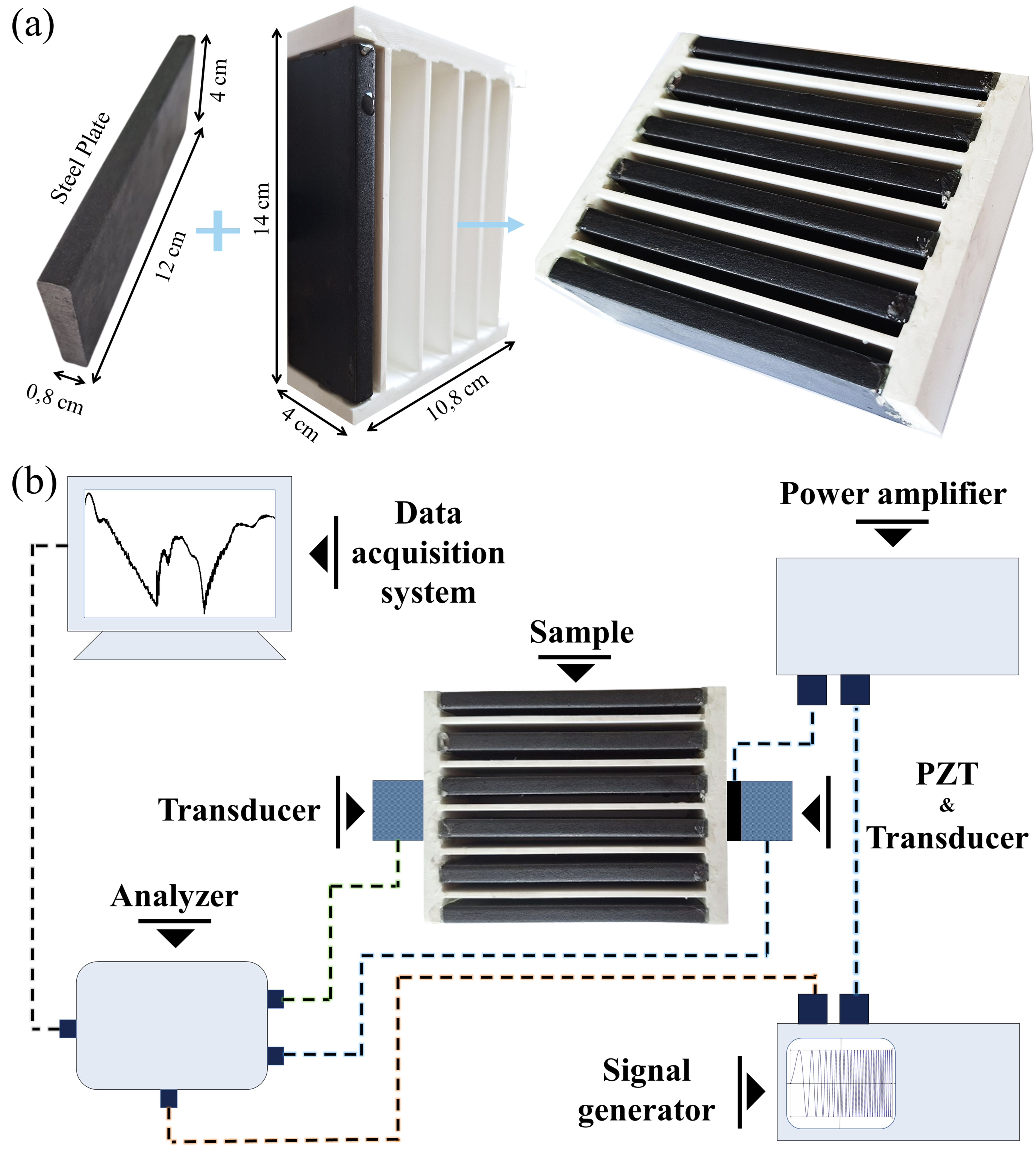}
    \caption{Illustrations of lab transmission measurement and sample fabrication. (a) the sample fabrication procedure, which involves printing a 3D structure using ABS filament and gluing steel plates using epoxy adhesive. (a) The principle of experimental transmission measurement comprises a signal generator, amplifier, transducers, analyzer, and data acquisition. The lab setup is showcased in \autoref{fig s1}.}
    \label{figure 2}
\end{figure}

To corroborate the analytical and numerical results, a prototype metasurface structure was fabricated by assembling Steel plates and an ABS structure fabricated by 3D printing with ABS filament, as shown in \autoref{figure 2}(a). The geometric parameters of the fabricated structure are $a =2$ \si{cm} and $l=12$ \si{cm}, for a height of $4$ \si{cm} and a homogeneous medium on both sides with a depth of $1$ \si{cm}.

\autoref{figure 2}(b) shows the experimental setup used to measure the elastodynamic response of the structure. We used a signal generator to generate a sweep signal connected to a power amplifier that can amplify the signal amplitude; the latter is connected to a PZT transmitter and a transducer to measure the input displacement. At the output, there is another transducer that can measure the received signal transmitted by the structure. In addition, this transducer is connected to an analyzer, which is linked to a laptop for data acquisition. As a result, transmitted and incident displacements were measured and recorded in order to calculate transmission curves. To better compare simulated and measured results, a 3D model was created in COMSOL, and the corresponding transmission curves were calculated. 

\section{Results and discussion}

\begin{figure}[!h]
   \centering
    \includegraphics[width=9 cm]{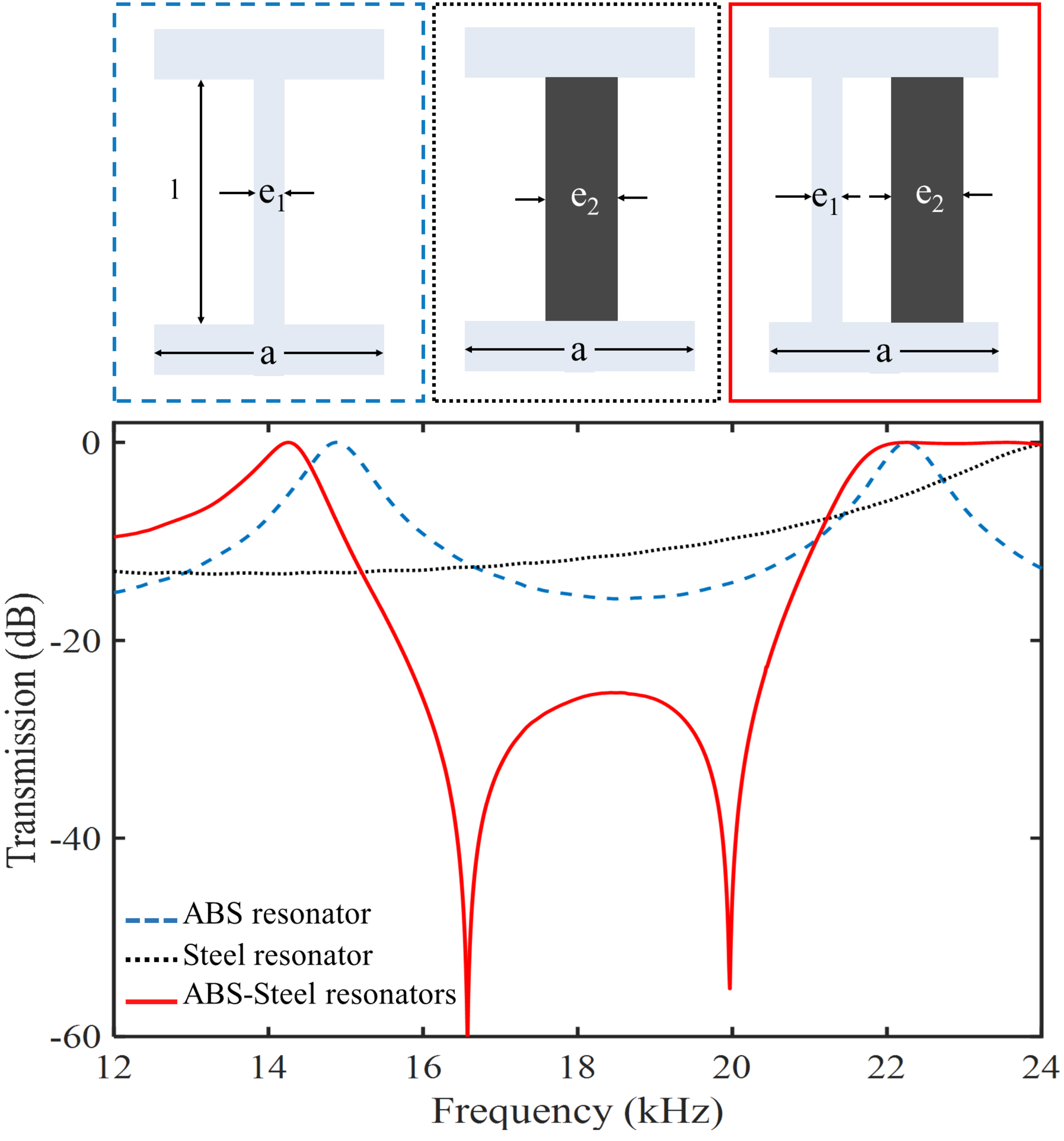}
    \caption{The mechanism behind the emergence of the w-shaped elastodynamic band gap. Transmission responses from the three unit cells. The dotted blue represents only the ABS resonator, the black dotted steel resonator, and the red line represents the coupling between both resonators.}
    \label{figure 3}
\end{figure}

Our study began with series of transmission simulations aimed at uncovering the mechanisms behind the formation of the W-shaped band gap in elastic plate metamaterials. We kept the periodicity constant in the y-axis fixed at $a= 2$ \si{cm} and used a plate length of $6a$, as shown in \autoref{figure 3}.
Initially, we constructed a simple Fabry-Perot resonator using ABS materials (represented by the dotted blue curve). This resulted in the emergence of a resonance peak with a Lorentzian profile at $n\lambda/2$ (where $\lambda$ represents the wavelength). This peak was depended to the material properties and was indicative of resonance frequencies.

Subsequently, we replaced the ABS material between the slits with steel plates (represented by the dotted black curve) to investigate the characteristics of Fabry-Perot resonators based on steel. Here, we observed that the resonance frequencies with Lorentzian profiles had shifted to higher frequencies, reaffirming their dependence on material properties.
The pivotal moment came when we combined both ABS and steel plates. This interaction between the two Fabry-Perot resonators, one constructed with ABS and the other with steel materials, gave rise to two distinct asymmetric Fano resonance peaks (represented by the solid red curve). These peaks were the result of coupling between the resonators and led to the formation of a broad W-band gap. This W-band gap was characterized by a width of $46\%$ and was centered around $16.6$ \si{kHz}. 
Importantly, the previously symmetric Fabry-Perot resonances had transformed into a series of asymmetrical resonances, indicating a Fano-like interactions in the system. This complex interplay of resonances and their coupling explained the emergence of the wide W-shaped band gap in the elastic plate metamaterial.

\begin{figure}[!h]
    \centering
    \includegraphics[width=9cm]{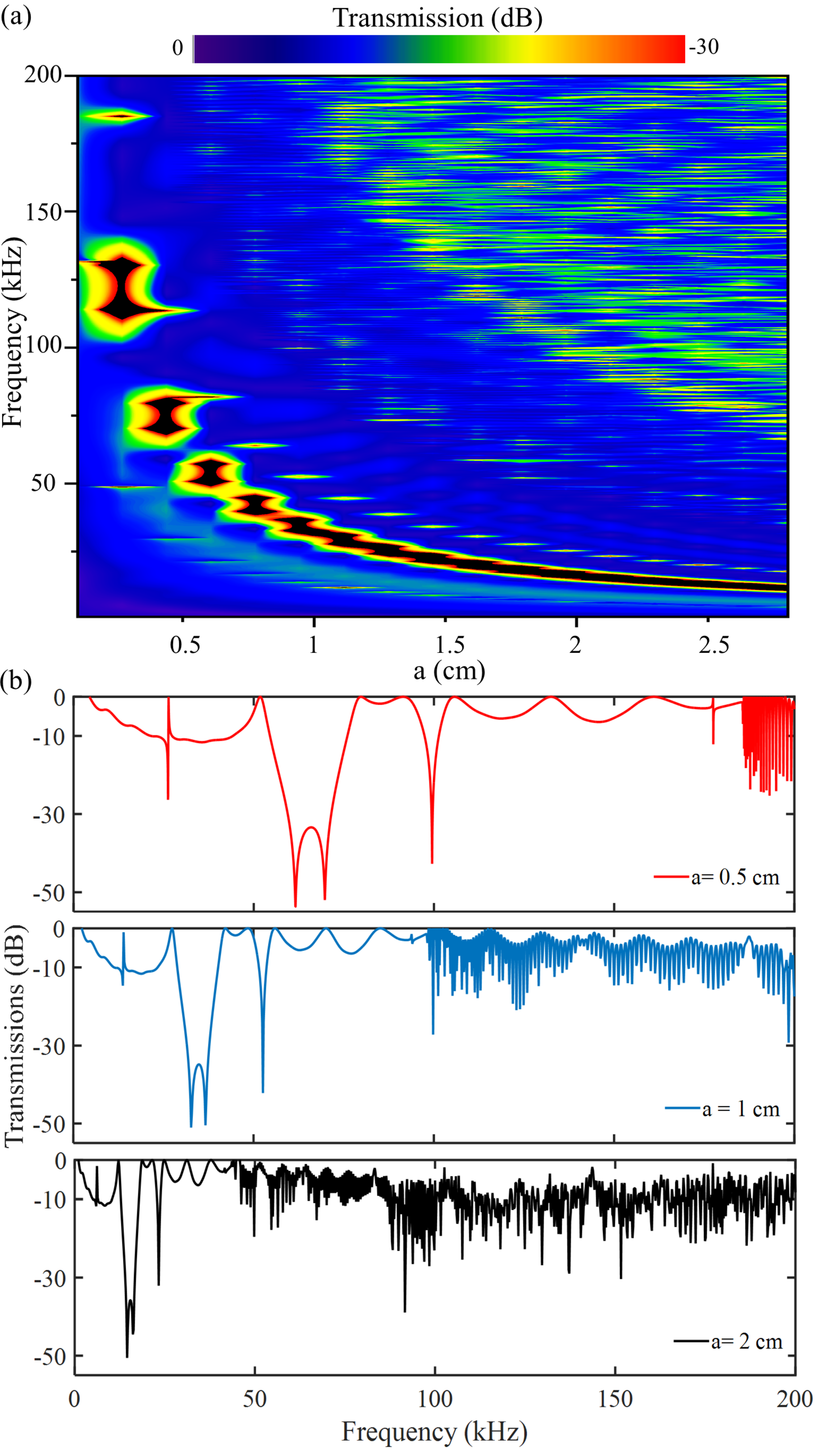}
    \caption{Parametric examination of the size effect on the w-shaped band gap. (a) Transmission map for the geometrical parameter $a$ scaling. (b) Transmission curves for three different geometrical parameter $a$ values.}
    \label{figure 4}
\end{figure}

After discussing the mechanism behind the formation of the W-shaped band gap, we now delve into an analysis of how various geometrical parameters affect this band gap. Specifically, we are examining the impact of the periodicity constant $a$, which ranges from $0.2$ to $2.7$ \si{cm}. The outcomes of this investigation are presented in \autoref{figure 4}(a), displaying a map of transmission as a function of both frequency and the parameter $a$.

Observations from the figure reveal that as the periodicity constant $a$ increases, the W-shaped band gap shifts towards lower frequencies. Furthermore, its width decreases compared to the initial scenario with $a$ set at $0.5$ \si{cm}. Another noteworthy phenomenon is the shifting of the diffraction limit, which is evident in \autoref{figure 4}(b).
In \autoref{figure 4}(b), the red curve represents the transmission of the unit cell when $a$ is equal to $0.5$ \si{cm}. Notably, the diffraction limit occurs at a frequency of $190$ \si{kHz} in this case. However, as $a$ increases, the diffraction limit shifts to lower frequencies, reaching $90$ \si{kHz} (blue curve) and eventually $50$ \si{kHz} (black curve) as $a$ becomes significantly larger. Conversely, when $a$ is smaller, the diffraction limit shifts to higher frequencies. It is important to emphasize that the diffraction limit is a phenomenon associated with the destructive interference of elastic waves, leading to distinctive propagation characteristics. In the supplementary materials, Figure 2 provides insight into the behavior of an array of elastic plate metamaterials at a fixed periodicity constant but at different frequencies, one above and the other after the diffraction limit. In the image above the diffraction limit, it is evident that the wavefronts propagate in parallel, indicating a plane longitudinal wave nature. However, after the diffraction limit, the wavefronts deviate from their plane wave nature and exhibit increased noise and complexity.

In summary, our analysis reveals that varying the periodicity constant $a$ significantly influences the frequency range of the W-shaped band gap, as well as the position of the diffraction limit, which is associated with changes in the nature of wavefront propagation.

\begin{figure}[!h]
    \centering
    \includegraphics[width=9cm]{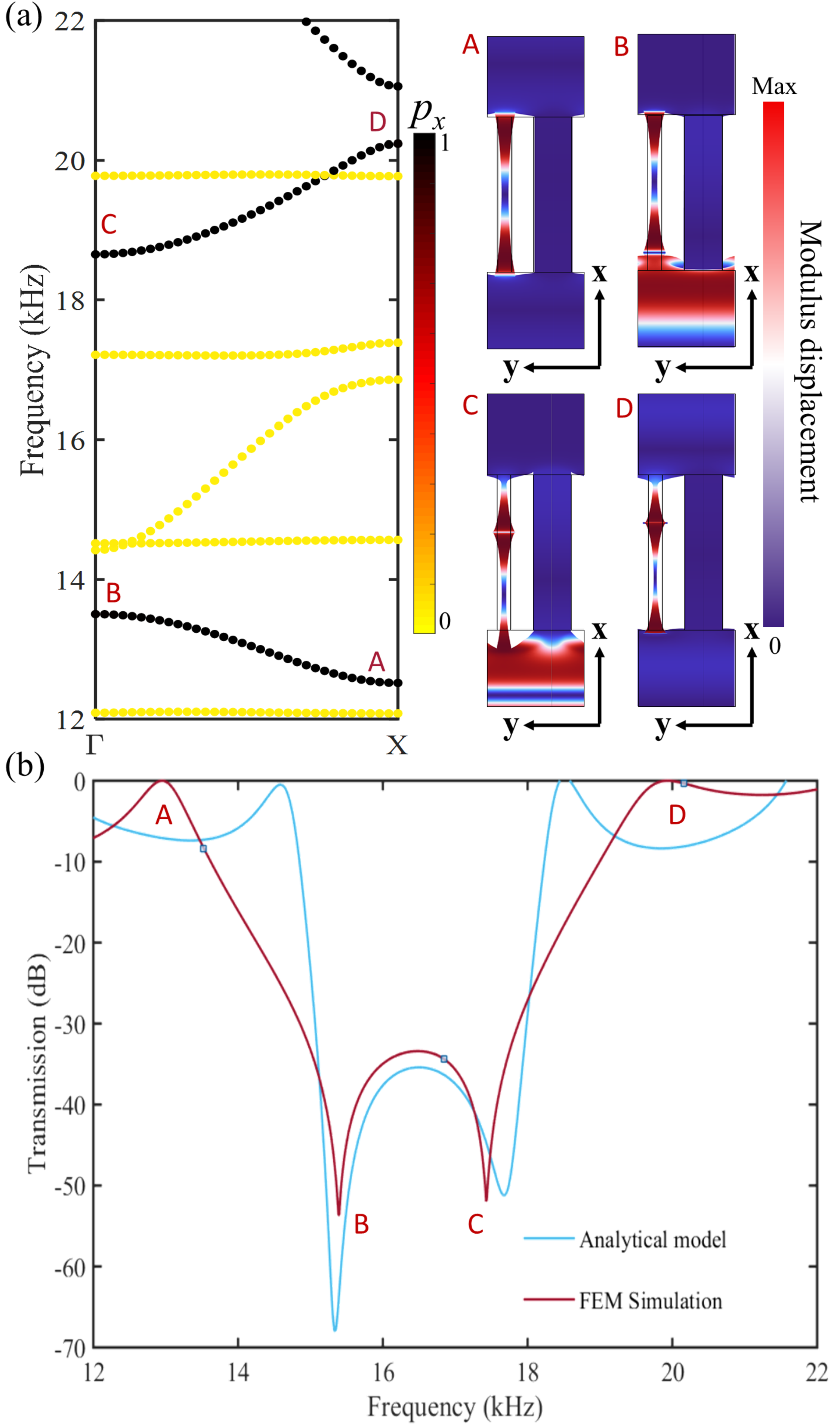}
    \caption{(a) Phononic dispersion of an infinite structure in the first Brillouin zone ($\Gamma X$) with geometrical parameters $a=0.5$ \si{cm} and $l=6a$, including screenshots of the modes appearing in the W-shaped band gap. The colorbar depicts the polarization of longitudinal modes, which appear in black, and transverse and flexural modes appear in yellow. (b) Analytical and FEM transmission curves for a finite structure with geometrical parameters $a=0.5$ \si{cm} and $l=6a$, with all modes existing in the w-shaped band gap.}
    \label{figure 6}
\end{figure}

\autoref{figure 6}(a) illustrates the phononic dispersion curve for the proposed unit cell, assuming periodicity in the direction of wave propagation, although our study does not involve such periodicity. We utilize these dispersion curves to elucidate the eigenmodes responsible for the emergence of the W-shaped band gap within the first Brillouin zone ($\Gamma X$).\\

It is evident that the longitudinal modes, denoted by black color, contrast with the modes polarized in the transverse plane. At the reducible wavenumber, specifically when it equals zero, points $B$ and $C$ appear at frequencies of $13.75$ \si{kHz} and $18.75$ \si{kHz}, respectively. These points correspond to the anti-resonance phenomena observed in the transmission analysis, exhibiting an attenuation of $-40$ \si{dB} at the same frequencies. Conversely, for the reducible wavenumber equal to $1$, points $A$ and $D$ are evident at frequencies of $12.5$ \si{kHz} and $20$ \si{kHz}, respectively. These points represent the resonance phenomena that result in complete transmission in the transmission analysis. To provide further insight, we have captured screenshots of these points, as depicted on the right side of \autoref{figure 6}(a). Points $A$, with coordinates ($k_x=1$, $f=12.5$ \si{kHz}), and $B$, with coordinates ($0$, $13.75$ \si{kHz}), signify the first Fano-like resonance induced by the two Fabry-Perot structures within the unit cell. Regarding the localization of total displacement in both screenshots at these points, it becomes apparent that the ABS plate undergoes a deformation involving dilatation and compression of the first order. On the other hand, points $C$, with coordinates ($k_x=0$, $f=18.75$ \si{kHz}), and $D$, with coordinates ($1$, $20$ \si{kHz}), represent the second Fano-like resonance generated by the two Fabry-Perot structures within the unit cell. Examining the total displacement localization in both screenshots at these points reveals that the ABS plate undergoes a deformation involving dilatation and compression of the second order.

The dispersion analysis demonstrates that the eigenmodes responsible for creating the w-shaped band gap pertain to infinite unit cells in the direction of longitudinal wave propagation. However, it is important to note that our objective is to achieve a band gap with just one unit cell in the direction of wave propagation, which diverges from the conditions assumed in the dispersion analysis.

The transmission spectrum for longitudinal waves, as determined through our analytical model and Finite Element Method (FEM) simulations, is presented in \autoref{figure 6}(b) for the purpose of comparison. It is readily apparent that the initial Fano-like peak, occurring at a frequency of $13.75$ \si{kHz}, comprises two symmetrical resonances. This symmetry is evident in the localization of total displacement. Similarly, the second Fano-like peak at $18.75$ \si{kHz} exhibits symmetry with respect to the first peak. Notably, both of these modes are situated within the ABS plate enclosed within the unit cell. Consequently, the deformation in this plate results from two distinct types of deformations, namely, compression and dilatation.

These resonance phenomena are accountable for the emergence of a w-shaped band gap, centered at a frequency of $16.25$ \si{kHz} and exhibiting a width of $46\%$, characterized by a substantial attenuation of $40$ \si{dB}. In the case of the analytical counterpart, the first peak appears at a frequency of $15$ \si{kHz} and is likewise composed of two symmetrical resonances. Subsequently, the second peak manifests itself at a frequency of $18$ \si{kHz}, giving rise to the w-shaped band gap within the analytical framework. It is pertinent to note that the w-shaped band gap, as determined analytically, also exhibits an attenuation of $40$ \si{dB}. In the context of normally incident longitudinal waves, the frequencies predicted by the analytical solution closely correspond to the numerical results, underscoring the correctness of our analytical assumptions and conditions.

\begin{figure}[!ht]
    \centering
    \includegraphics[width=9cm]{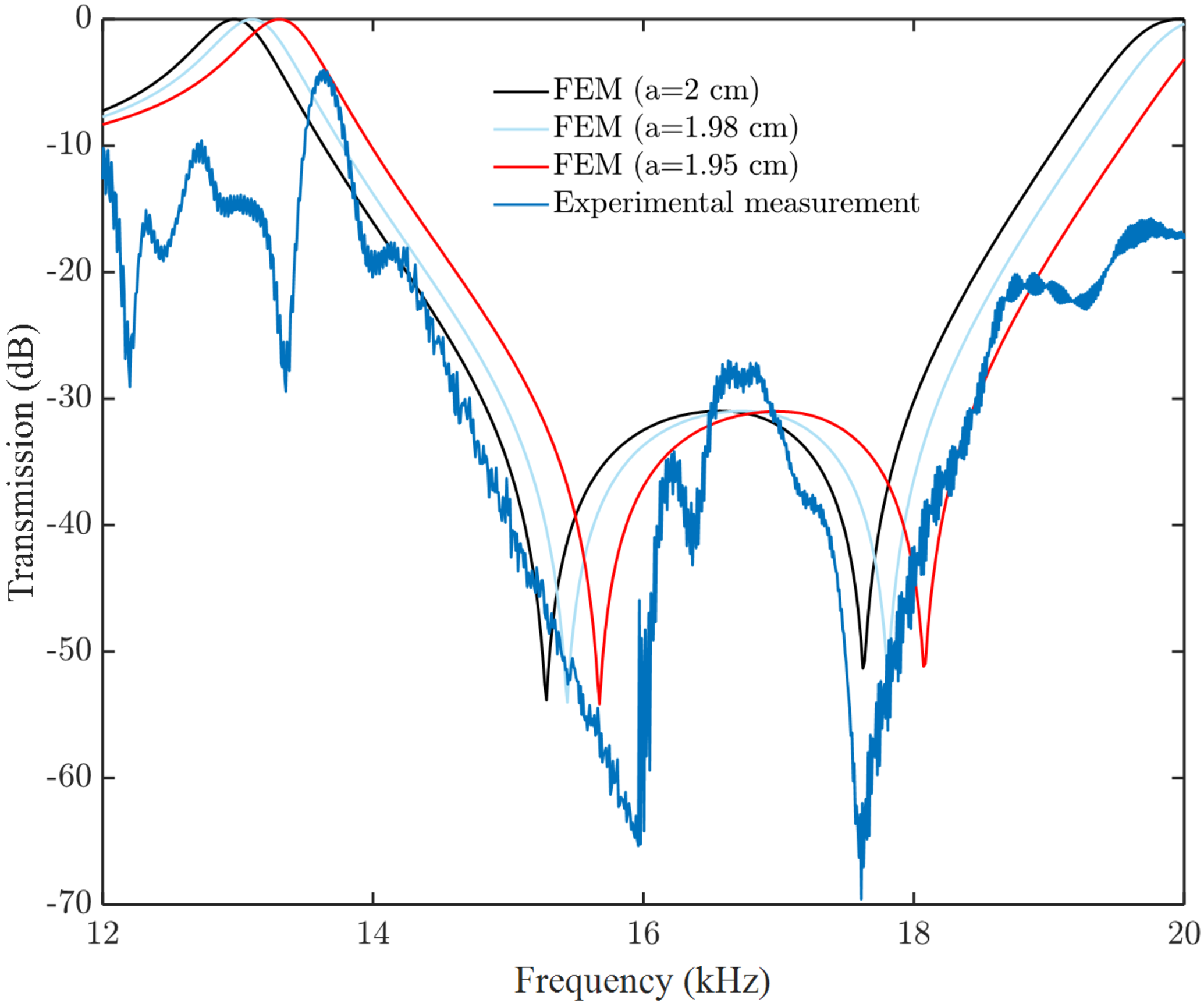}
    \caption{Experimental curve for composite elastic metasurface with $a=2$ \si{cm}, compared with FEM transmission curves for the metasurface structure with geometric parameters $a = 2$ \si{cm}, $a = 1.98$ \si{cm} and $a = 1.95$ \si{cm}.}
    \label{figure 7}
\end{figure}

As seen in \autoref{figure 4}(a), the W-shaped band gap is scalable, which indicates that the same phenomenon can occur at low frequencies corresponding to the higher value of the constant parameter $a$. In order to accomplish this, we computed the numerical W-shaped band for $a=2$ \si{cm} and then manufactured a sample of the same size. \autoref{figure 7} shows the comparison of this investigation. The measured transmission curve shows the two Fano resonances: the first has a resonance peak at $13.75$ \si{kHz} and an antiresonance at $16.25$ \si{kHz}. The second Fano resonance has an antiresonance of $18$ \si{kHz} and a resonance peak of $20$ \si{kHz}. The experimental results match with their computational equivalent, which shows the presence of two Fano resonances at this range of frequencies for $a=2$ \si{cm}. But there is a shift between the numerical and experimental transmissions, most likely due to material qualities; in the calculations, all materials were assumed to be isotropic; nevertheless, this was not the case in the experiment. However, we can see that the W-shaped band gap arises in the same frequency range, and both Fano resonances are valid. That means that the W-shaped band gap is generated by the interaction of two Fabry-Perots, one made of ABS and the other of steel.
Upon comparing the results obtained from experimental measurements and FEM simulations with those from the analytical model, it becomes evident that the formation of the w-shaped band gap results from the coupling between two Fabry-Perot resonators formed by the rectangular slits positioned between the steel and ABS materials. Furthermore, the analytical model proposed in this study aligns remarkably well with the FEM simulations and the experiment results, demonstrating that the broad w-shaped band gap is not only evident in terms of matching transmission peak frequencies but also in terms of attenuation levels.

\section{Conclusion}
To conclude, we have successfully demonstrated a novel solution to address the limitations inherent in traditional large-scale elastic metamaterials. Our approach involves the creation of an elastic composite metasurface, utilizing a specialized unit cell design capable of generating a distinctive W-shaped bandgap at lower frequencies. This innovative structure comprises dual Fabry-Perot resonators crafted from steel and ABS materials. The intricate interplay between the longitudinal modes and localized modes within these resonators gives rise to the unique W-shaped bandgap phenomenon.

Maxwell stress tensor distribution calculated by using an optical lens that focuses the laser beam on the mass-spring system. (a) The Maweel stress tensor's x, y, and z components. (b) The Maxwell Stress Tensor's modulus.

Finite mass-spring chain optically trapped in the temporal domain. (a) The calculated phononic dispersion of a finite mass-spring chain under optical traps for $40$ unit cells along the $x$-axis. The modulus $|U_x|$ of the complex response function is plotted on a false-color scale versus the $x$-component of the wave vector and frequency. The black solid line represents the numerical computation of dispersion for an infinite mass-spring chain. (b) The calculated transmission in dB between the modulus of the input $x$-component displacement and the modulus of the output  after $40$ unit cells $x$-component displacement.

\section*{Acknowledgment}
The authors acknowledge the support of the ANR PNanoBot project (contract "ANR-21-CE33-0015").\\
This work is funded by the European Union, under Marie Skłodowska-Curie Actions Postdoctoral Fellowships (Project 101149710—STM).\\
K.K.D. acknowledges the support of the Polish National Science Centre (NCN) in the form of the grant awarded as a part of the SONATA 18 program, project No. 2022/47/D/ST5/00280 under the name ”Active micro-scale mechanical metamaterials controlled by the external stimuli”. 
This work was partially supported by a program of the Polish Ministry of Science under the title ‘Regional Excellence Initiative’, project no. RID/SP/0050/2024/1.
\appendix

\section{ }
\begin{figure}[!h]
    \centering
    \includegraphics[width=1\linewidth]{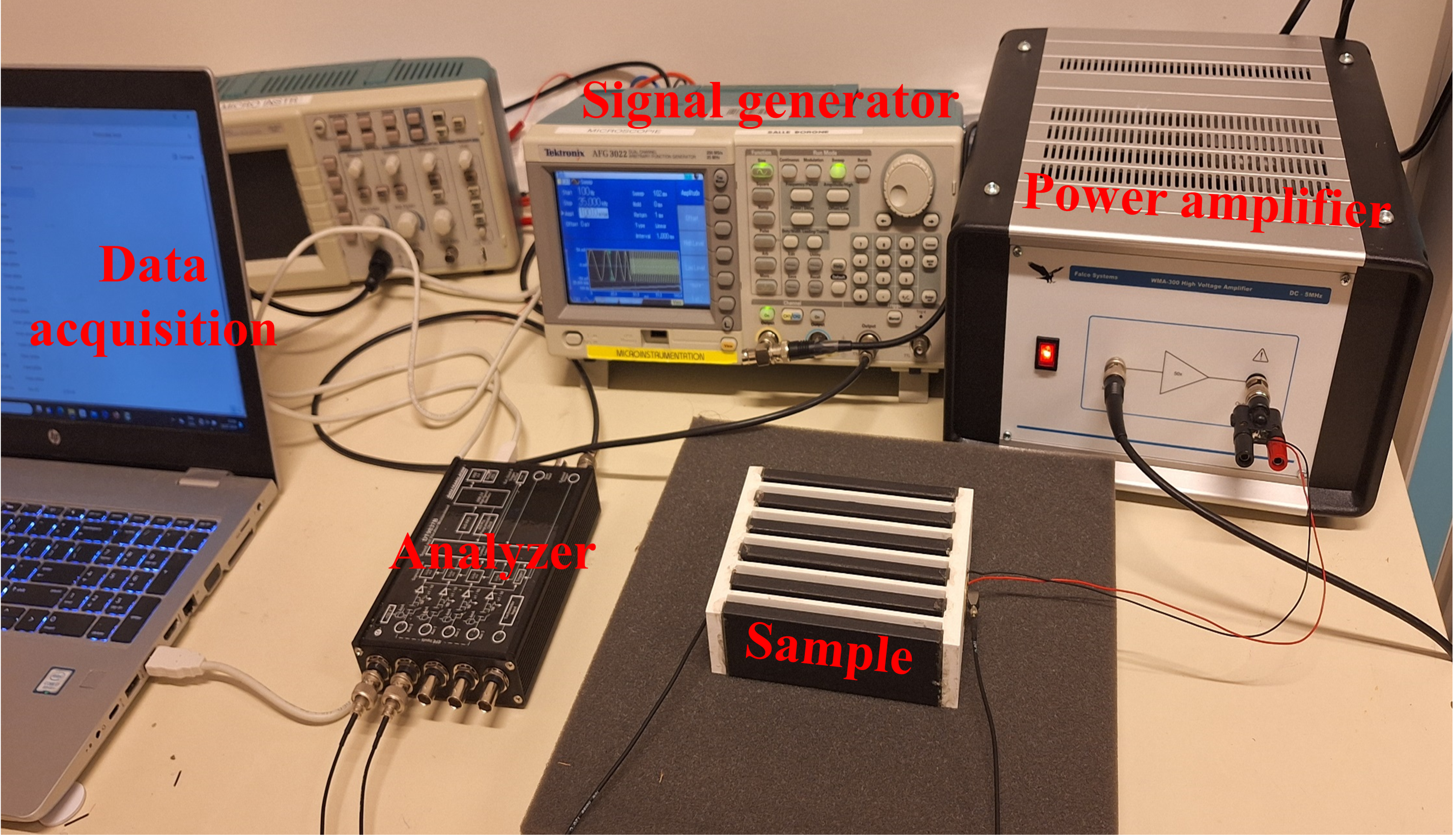}
    \caption{Experimental transmission measurement setup consisting of a signal generator, amplifier, transducers, analyzer, and data acquisition.}
    \label{fig s1}
\end{figure}

\begin{figure}[!h]
    \centering
    \includegraphics[width=1\linewidth]{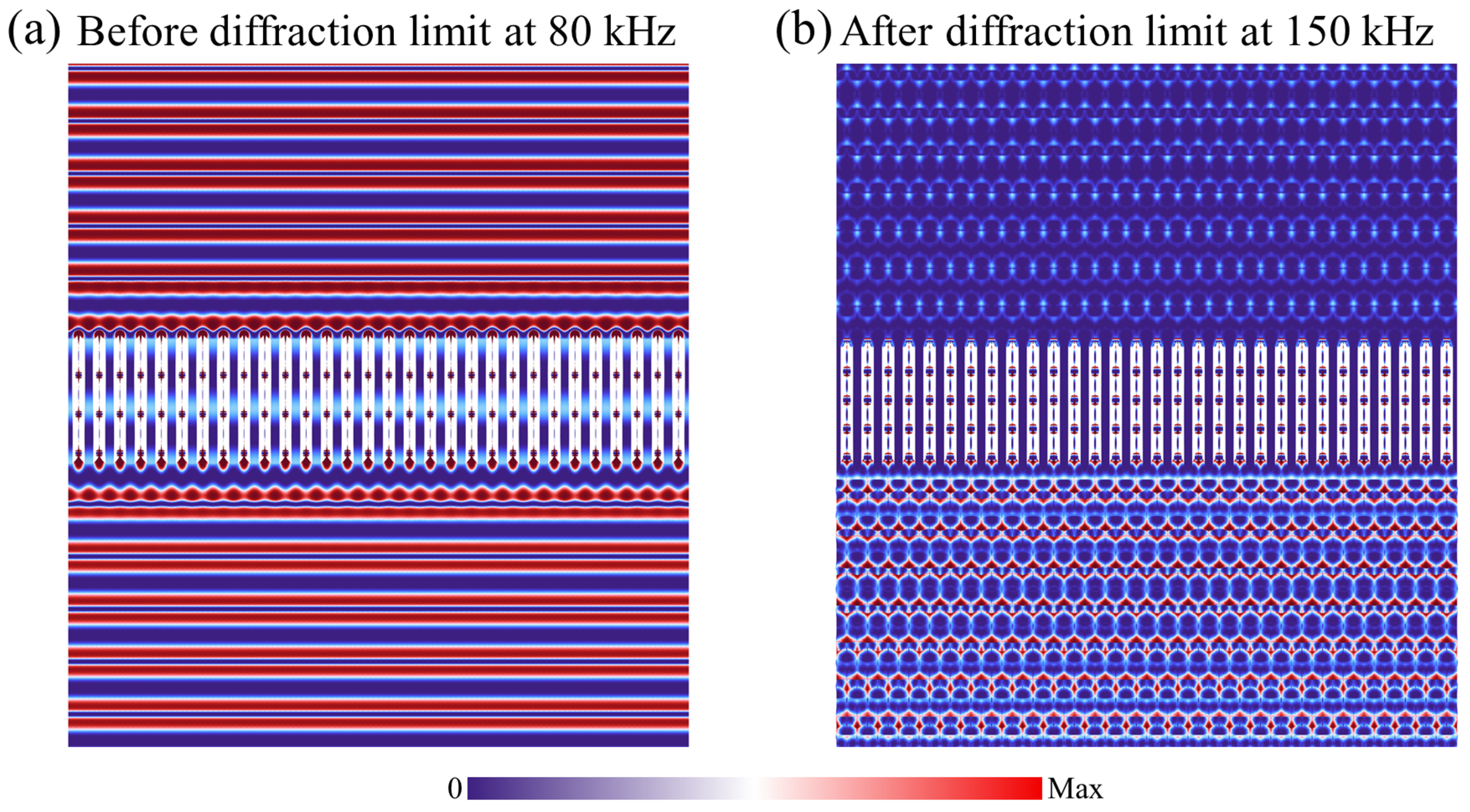}
    \caption{Illustrations of wave propagation assuming the geometrical parameter $a=1$ \si{cm}. (a) At a frequency of $100$ \si{kHz}, before the diffraction limit. (b) After reaching the diffraction limit at $150$ \si{kHz}.}
    \label{fig s2}
\end{figure}

\begin{figure}[!h]
    \centering
\includegraphics[width=1\linewidth]{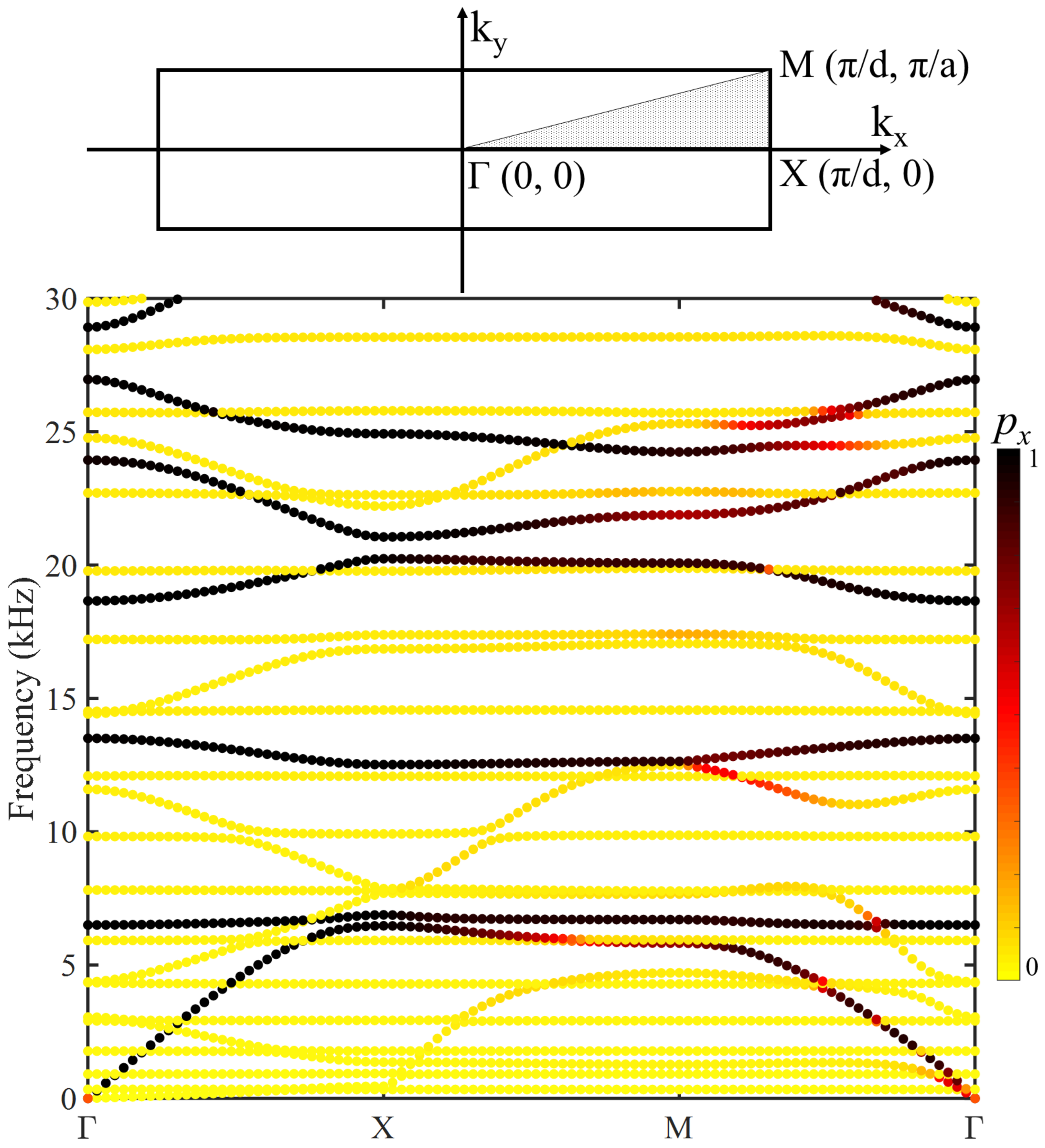}
    \caption{Full phononic dispersion of an infinite pattern in the first irreducible Brillouin zone ($\Gamma \rm X \rm M \Gamma$).}
    \label{fig s3}
\end{figure}

\bibliography{Mybib}

\end{document}